\documentclass[11pt,english]{article}
\usepackage[T1]{fontenc}
\usepackage[utf8]{inputenc}
\usepackage{geometry}
\usepackage{mathrsfs}
\usepackage{amsmath}
\usepackage{amssymb}
\usepackage{graphicx}
\usepackage{setspace}
\usepackage{csquotes}
\makeatletter
\newcommand{\lyxaddress}[1]{
	\par {\raggedright #1
	\vspace{1.4em}
	\noindent\par}
}

\makeatother
\usepackage{times}
\usepackage{babel}
\usepackage[sorting=none, maxbibnames=99]{biblatex}
\begin{document}
\global\long\def\ub{\boldsymbol{u}}%

\global\long\def\gradu{\nabla\ub}%

\global\long\def\cb{\boldsymbol{c}}%

\global\long\def\sb{\boldsymbol{\sigma}}%

\global\long\def\db{\boldsymbol{\delta}}%

\global\long\def\gdot{\dot{\gamma}}%

\global\long\def\zeq{Z_{\text{eq}}}%

\global\long\def\wi{\text{Wi}}%

\global\long\def\wmax{\text{Wi}_{\text{max}}}%

\title{Improving the functionality of non-stretching approximations}
\author{Vickie Chen, Brandon Wang, Joseph D. Peterson}
\maketitle

\lyxaddress{Department of Chemical and Biomolecular Engineering, University of
California, Los Angeles, 420 Westwood Plaza, Los Angeles CA 90095}
\begin{abstract}
Entangled polymers are an important class of materials for their toughness,
processability, and functionalizability. However, physically detailed
modeling of highly entangled polymers can prove challenging, especially
as one considers additional layers of physical or chemical complexity.
To address these challenges, we present a series of generalizations
for the useful "non-stretching" approximation, using asymptotic
methods to formalize and expand the analysis. First, we rederive the
popular non-stretching Rolie Poly model and extend it second order,
reintroducing effects from finite chain stretching. Then, we extended
the non-stretching framework to other special cases, accounting for
flow-induced disentanglement, polydispersity, and reversible scission
reactions. Benchmark calculations confirm that non-stretching models
derived via systematic asymptotic methods provide excellent and improvable
approximations for the rheology of well-entangled polymer constitutive
equations with finite-time stretch relaxation dynamics.
\end{abstract}

\section{\protect\label{sec:Introduction}Introduction}

From commodity thermoplastics to biofilms and shampoos, entangled
polymers are an industrially significant and physically interesting
class of materials to study. Compared to unentangled polymers, entangled
polymers have higher elasticity, higher viscosity, and greater toughness
- properties that can be modified by strong flow or tuned by plasticizers,
chemical additives, or changes in size/polydispersity/architecture
of the polymers themselves. With every added layer of physical detail
or chemical functionality, however, interpreting experimental observations
and constructing new predictive models becomes increasing challenging. 

To account for new physics in an entangled polymer system, it is useful
to build upon an accepted "chassis" that has been validated for
simpler related problems. For example, the Rolie Poly model is a popular
and reliable model for the rheology of well-entangled, linear, monodisperse
polymer melts \cite{likhtman2003simple}, and it has been adapted
to study the effect of added solvent/plasticizers \cite{peterson2016shear,peterson2019does},
polymerizatoin reactions \cite{peterson2021predictions,peterson2021constitutive},
flow-induced disentanglement \cite{dolata2023thermodynamically},
polydispersity \cite{boudara2019nonlinear}, finite extensibility
\cite{kabanemi2009nonequilibrium} and more. Surprisingly, however,
a useful simplification to the "chassis" model - introduced in
the original Rolie Poly paper - has not yet been translated to any
of its successor variants as far as we know.

The "non-stretching" approximation is a useful simplification
for understanding the nonlinear rheology of many well-entangled polymer
systems. When highly entangled polymers relax their configuration/stress,
they do so subject to the topological constraints of their entanglements
with surrounding chains. Tube theory, conceptualized by de Gennes
\cite{de1971reptation} and refined by Doi and Edwards \cite{doi1978dynamics,doi1988theory}
suggests that the net effect of entanglements is to constrain polymer
relaxation to occur within a "tube" that constrains the lateral
motion of a polymer but not its curvilinear motion. As a result of
this confining tube, entangled polymers (unlike unentangled polymers)
have an orientational relaxation time that is significantly delayed
compared to their stretch relaxation time. For deformations occurring
on timescales comparable to the orientational relaxation time, chain
stretch is only weakly perturbed from equilibrium and - to leading
order - chains can be treated as "non stretching" \cite{likhtman2003simple}.

The non-stretching approximation leads to a universality class for
highly entangled polymers - as long as chains are not stretching,
the actual details of the entanglement number $Z$ do not matter to
leading order. This simplifies the design of computational studies
and the interpretation of experimental data by eliminating an unknown/uncertain/irrelevant
degree of freedom from the analysis. The non-stretching approximation
was successfully employed in the original non-linear constitutive
equation by Doi and Edwards \cite{doi1978dynamics}, the full-chain
model by Milner McLeish and Likhtman \cite{graham2003microscopic},
and the Rolie Poly model \cite{likhtman2003simple}. However, the
non-stretching approximation has not yet been extended to more recent
variants of the Rolie Poly model.

There are, of course, admitted weaknesses to the non-stretching approximation.
The non-stretching approximation is most accurate for very large entanglement
numbers, $Z\gg100$, which are difficult to reproduce experimentally
(at least for monodisperse systems). Additionally, the non-stretching
approximation is a singular limit of the dynamics, and chain stretching
must always be possible if the rate of deformation becomes fast enough.
In extreme cases, the omission of chain stretching can even lead to
predictions of flows that reach infinite shear rates finite time \cite{chen2024finite},
and some means of systematically reintroducing chain stretching is
required.

In the present manuscript, we will provide a systematic review of
the non-stretching approximation in the Rolie Poly model (section
\ref{sec:nRP2}). We will derive the original non-stretching approximation
as a singular limit of the full Rolie Poly model in the limit of $1/Z\ll1$,
and we will extend this derivation to a second-order accuracy variant
that solves the aforementioned problem of finite-time blow-up. We
will then repeat the analysis for leading-order terms in variants
of the Rolie Poly model that account for flow-induced disentanglement
(section \ref{sec:nDO}) , polydisperse molecular weight distributions
(section \ref{sec:nRDP}), and living polymer reactions (section \ref{sec:nLRP}).
Our goal is to give the reader a more complete understanding for the
origins and generalizability of the non-stretching approximation in
entangled polymer systems. 

\section{\protect\label{sec:nRP2}Improved Accuracy}

\subsection{\protect\label{sec:Equations-and-Nondimensionalizat}Starting Equations}

The Rolie Poly model is a single-mode approximation of the more complete
GLaMM model for monodisperse well entangled linear polymers \cite{likhtman2003simple}.
The Rolie Poly model accounts for affine deformation in flow and stress
relaxation by reptation, chain retraction, and convective constraint
release (CCR), relating the fluid stress $\sb'$ to a polymer configuration
tensor $\cb$:

\begin{equation}
\frac{\partial\cb}{\partial t'}+\ub'\cdot\nabla\boldsymbol{c}-\gradu'{}^{T}\cdot\boldsymbol{c}-\boldsymbol{c}\cdot\gradu'=-\frac{1}{\tau_{D}}(\boldsymbol{c}-\db)-\frac{2}{\tau_{R}}(1-\frac{1}{\lambda})(\boldsymbol{c}+\beta\lambda^{2\delta}(\boldsymbol{c}-\db))\label{eq:Original_Rolie_Poly}
\end{equation}

\begin{equation}
\lambda=\sqrt{\frac{1}{3}\text{tr}\cb}
\end{equation}

\begin{equation}
\sb'=G_{e}(\cb-\db)
\end{equation}

where $\ub'$ is the velocity field, $t'$ is the time, $\lambda$
is the chain stretch, $G_{e}$ is the shear modulus, $\tau_{D}$ is
the reptation time, $\tau_{R}$ is the stretch relaxation time, and
$\db$ is the identity tensor. The parameters $\beta$ and $\delta$
govern the effectiveness of CCR and its sensitivity to stretching,
respectively. The original Rolie Poly paper recommends $\beta=1$
and $\delta=-1/2$ for a good fit to experimental data in non-reversing
flows, but any fixed value of $\beta>0$ has been shown to yield non-physical
predictions in reversing flows \cite{drucker2024definite}. Alternative
proposals for $\beta$ that change sign during reversals have been
suggested \cite{ianniruberto2001simple,drucker2024definite}, so our
analysis will allow $\beta$ to be a general function of the material
parameters and chain configuration.

To nondimensionalize, we assume a characteristic velocity $U_{c}$,
stress $G_{e}$, lengthscale $H$, and timescale $\tau_{D}$ for the
flow of interest. We define dimensionless variables by rescaling against
these characteristic values, using $x,y,z$ as underlying coordinate
axis:
\begin{equation}
\ub=\frac{\ub'}{U_{c}}\hspace{1cm}x=\frac{x'}{H}\hspace{1cm}t=\frac{t'}{\tau_{D}}\hspace{1cm}\sb=\frac{\sb'}{G_{e}}
\end{equation}
Substituting the dimensionless variables into equation \ref{eq:Original_Rolie_Poly},
we obtain:

\begin{equation}
\partial_{t}\boldsymbol{c}+\left(\ub\cdot\nabla\boldsymbol{c}-\nabla\ub{}^{T}\cdot\boldsymbol{c}-\boldsymbol{c}\cdot\nabla\ub\right)\text{\ensuremath{\left[\frac{U_{c}\tau_{D}}{H}\right]}}=-(\boldsymbol{c}-\db)-6\left[\frac{\tau_{D}}{3\tau_{R}}\right](1-\frac{1}{\lambda})(\boldsymbol{c}+\frac{\beta}{\lambda}(\boldsymbol{c}-\db))
\end{equation}

\begin{equation}
\lambda=\sqrt{\frac{1}{3}\text{tr}\cb}
\end{equation}

\begin{equation}
\sb=\cb-\db
\end{equation}

The terms in square brackets are dimensionless numbers, which we identity
as the Weissenberg number, $\wi=U_{c}\tau_{D}/H$, and the entanglement
number, $Z=\tau_{D}/3\tau_{R}$, respectively. The relationship between
the entanglement number $Z$ and the reptation/stretch relaxation
times is valid for very large entanglement numbers, which are the
focus of our present study. To simplify the notation in future equations,
we define $\overset{\nabla}{\cb}=\partial_{t}\boldsymbol{c}+\left(\ub\cdot\nabla\boldsymbol{c}-\gradu{}^{T}\cdot\boldsymbol{c}-\boldsymbol{c}\cdot\gradu\right)\text{Wi}$
as the upper-convected time derivative of the configuration tensor.

The full non-dimensionalized Rolie-Poly model is given by:
\begin{equation}
\overset{\nabla}{\boldsymbol{c}}=-(\boldsymbol{c}-\db)-6Z(1-\frac{1}{\lambda})(\boldsymbol{c}+\frac{\beta}{\lambda}(\boldsymbol{c}-\db))\label{eq:full_RP_ndim}
\end{equation}

\begin{equation}
\lambda=\sqrt{\frac{1}{3}\text{tr}\cb}
\end{equation}

\begin{equation}
\sb=\cb-\db
\end{equation}

A non-stretching approximation for the Rolie Poly model was put forward
in the original Rolie Poly paper \cite{likhtman2003simple}. Using
the same non-dimesionalization scheme, the non-stretching Rolie Poly
(nRP) model is given by:

\begin{equation}
\overset{\nabla}{\boldsymbol{c}}=-(\boldsymbol{c}-\delta)-\frac{2}{3}(\cb:\gradu)(\boldsymbol{c}+\beta(\boldsymbol{c}-\db))\label{eq:full_nRP_ndim}
\end{equation}

\begin{equation}
\text{tr}\cb=3
\end{equation}

\begin{equation}
\sb=\cb-\db
\end{equation}

The nRP model provides a useful approximation of the RP model in flows
with $Z\gg1$ and Wi $\ll Z$, but for a more accurate model (and
one that avoids finite-time blow-up \cite{chen2024finite}) we will
extend the non-stretching approximation to higher order.

\subsection{\protect\label{sec:Derivation-of-Asymptotic}Derivation}

\subsubsection{\protect\label{subsec:Setting-up-Asymptotic}Setting up Asymptotic
Expansion}

As summarized in section \ref{sec:Introduction}, a non-stretching
approximation is appropriate for highly entangled polymers $Z\gg1$
under flow conditions where the chain's stretch is not very far from
its equilibrium value, $|\lambda-1|\ll1$. In general, this is expected
to occur when the strain rate satisfies $\wi\ll Z$ and $Z\gg1$.

These asymptotic limits can be formally defined as a fixed $\wi$
and an entanglement number $Z$ that goes to infinity - or equivalently
an inverse entanglement number $\epsilon=1/Z$ that is vanishingly
small, $\epsilon\ll1$. As an ansatz, we will assume that a regular
perturbation expansion exists in the limit of $\epsilon\ll1$ such
that:

\begin{align}
\boldsymbol{c}(t,x,\text{Wi},Z) & =\boldsymbol{c}_{0}(t,x,\text{Wi})+\epsilon\boldsymbol{c}_{1}(t,x,\text{Wi})+\epsilon^{2}\boldsymbol{c}_{2}(t,x,\text{Wi})+\cdots\label{eq:c_expansion}
\end{align}

\begin{equation}
\frac{1}{\lambda}=a=a_{0}+\epsilon a_{1}+\epsilon^{2}a_{2}+\cdots\label{eq:a_expansion}
\end{equation}

\begin{equation}
\beta=\beta_{0}+\epsilon\beta_{1}+\epsilon^{2}\beta_{2}+\cdots\label{eq:beta_expansion}
\end{equation}

With the expansion for \textbf{c} and $\lambda=\sqrt{\text{tr}(\boldsymbol{c})/3}$,
we can write $a=1/\lambda$ as

\begin{equation}
a=\frac{1}{\lambda}=\left[\frac{\text{\text{tr}\ensuremath{\left(\boldsymbol{c}_{0}+\epsilon\boldsymbol{c}_{1}+\epsilon^{2}\boldsymbol{c}_{2}\right)}}}{3}\right]^{-1/2}=\left[\frac{\text{\text{tr}\ensuremath{\left(\boldsymbol{c}_{0}\right)}}}{3}\left(1+\epsilon\frac{\text{\text{tr}\ensuremath{\left(\boldsymbol{c}_{1}\right)}}}{\text{\text{tr}\ensuremath{\left(\boldsymbol{c}_{0}\right)}}}+\epsilon^{2}\frac{\text{\text{tr}\ensuremath{\left(\boldsymbol{c}_{2}\right)}}}{\text{\text{tr}\ensuremath{\left(\boldsymbol{c}_{0}\right)}}}\right)\right]^{-1/2}\label{eq:a_expansion-1}
\end{equation}

We can use a Taylor Series to expand equation \ref{eq:a_expansion-1}
about $\epsilon=0$:

\begin{equation}
a=a_{0}+\epsilon a_{1}+\epsilon^{2}a_{2}\approx\left[\frac{3}{\text{tr}(\boldsymbol{c}_{0})}\right]^{1/2}\left[1-\epsilon\frac{1}{2}\frac{\text{tr}(\boldsymbol{c}_{1})}{\text{tr}(\boldsymbol{c}_{0})}+\epsilon^{2}\left(\frac{3}{8}\left[\frac{\text{tr}(\boldsymbol{c}_{1})}{\text{tr}(\boldsymbol{c}_{0})}\right]^{2}-\frac{1}{2}\frac{\text{tr}(\boldsymbol{c}_{2})}{\text{tr}(\boldsymbol{c}_{0})}\right)\right]
\end{equation}

Therefore, we have:

\begin{equation}
a_{0}=\left[\frac{3}{\text{tr}(\boldsymbol{c}_{0})}\right]^{1/2}\label{eq:a0_defn}
\end{equation}

\begin{equation}
a_{1}=-\frac{1}{2}a_{0}\frac{\text{tr}(\boldsymbol{c}_{1})}{\text{tr}(\boldsymbol{c}_{0})}\label{eq:a1_defn}
\end{equation}

\begin{equation}
a_{2}=a_{0}\left(\frac{3}{8}\left[\frac{\text{tr}(\boldsymbol{c}_{1})}{\text{tr}(\boldsymbol{c}_{0})}\right]^{2}-\frac{1}{2}\frac{\text{tr}(\boldsymbol{c}_{2})}{\text{tr}(\boldsymbol{c}_{0})}\right)
\end{equation}

Combining the expansions for $\cb$, a, and $\beta$ with equation
\ref{eq:full_RP_ndim}, the Rolie-Poly model can be written as

\[
\overset{\nabla}{\boldsymbol{c}}_{0}+\epsilon\overset{\nabla}{\boldsymbol{c}}_{1}+\epsilon{}^{2}\overset{\nabla}{\boldsymbol{c}}_{2}=-(\boldsymbol{c}_{0}+\epsilon\boldsymbol{c}_{1}+\epsilon{}^{2}\boldsymbol{c}_{2}-\delta)-6\frac{1}{\epsilon}(1-a_{0}-\epsilon a_{1}-\epsilon^{2}a_{2})(\boldsymbol{c}_{0}+\epsilon\boldsymbol{c}_{1}+\cdots
\]

\begin{equation}
\epsilon{}^{2}\boldsymbol{c}_{2}+(\beta_{0}+\epsilon\beta_{1}+\epsilon^{2}\beta_{2})(a_{0}+\epsilon a_{1}+\epsilon^{2}a_{2})(\boldsymbol{c}_{0}+\epsilon\boldsymbol{c}_{1}+\epsilon{}^{2}\boldsymbol{c}_{2}-\delta))+\mathscr{O}(\epsilon^{3})\label{eq:expanded_nRP2}
\end{equation}

\subsubsection{\protect\label{subsec:Leading-order-solution}Leading order solution}

We rearrange equation \ref{eq:expanded_nRP2} as a series expansion
in terms of the small parameter $\epsilon$. Each term in this series
is multiplied by a different power of $\epsilon$, starting from the
leading order $\epsilon^{-1}$, followed by $\epsilon^{0}$, and continuing
with higher powers of $\epsilon$. \ref{eq:expanded_nRP2} now looks
like:

\begin{equation}
\epsilon^{-1}\left[6\left(1-a_{0}\right)\left(\boldsymbol{c}_{0}+\beta_{0}a_{0}\left(\boldsymbol{c}_{0}-\delta\right)\right)\right]+\epsilon^{0}\left[\overset{\nabla}{\boldsymbol{c}}_{0}+\left(\boldsymbol{c}_{0}-\delta\right)+\cdots\right]+\epsilon^{1}\left[\overset{\nabla}{\boldsymbol{c}}_{1}+\boldsymbol{c}_{1}+\cdots\right]+\cdots=0\label{eq:regrouped}
\end{equation}

To satisfy equation \ref{eq:regrouped} in the limit of $\epsilon\to0$,
every grouping of terms at each order in $\epsilon$ must independently
be equal to zero. Starting from the $\mathscr{O}(\epsilon^{-1})$
terms, we have:

\begin{equation}
6(1-a_{0})(\boldsymbol{c}_{0}+\beta_{0}a_{0}(\boldsymbol{c}_{0}-\delta))=0
\end{equation}

which is satisfied by $a_{0}=1$. Note that physically this means
that the chain stretch is at its equilibrium value; with $a_{0}=1$
and equation \ref{eq:a0_defn}, we can obtain $\text{tr}(\boldsymbol{c}_{0})$.

\begin{equation}
\text{tr}(\boldsymbol{c}_{0})=3\label{eq:tr(c0)}
\end{equation}

Next, we proceed to the $\mathscr{O}(\epsilon^{0})$ terms in equation
\ref{eq:regrouped}:

\[
\partial_{t}\boldsymbol{c}_{0}+(\ub\cdot\nabla\boldsymbol{c}_{0}-\gradu{}^{T}\cdot\boldsymbol{c}_{0}-\boldsymbol{c}_{0}\cdot\gradu)\text{Wi}+(\boldsymbol{c}_{0}-\boldsymbol{\delta})+\cdots
\]

\begin{equation}
+6(1-a_{0})(\boldsymbol{c}_{1}+\beta_{1}a_{0}(\boldsymbol{c}_{0}-\boldsymbol{\delta})+\beta_{0}a_{1}(\boldsymbol{c}_{0}-\boldsymbol{\delta})+\beta_{0}a_{0}\boldsymbol{c}_{1})+6(-a_{1})(\boldsymbol{c}_{0}+\beta_{0}a_{0}(\boldsymbol{c}_{0}-\boldsymbol{\delta}))=0\label{eq:c0_eqn}
\end{equation}

Equation \ref{eq:c0_eqn} requires knowing information from the first
correction terms, namely $a_{1}$. Fortunately, from equation \ref{eq:tr(c0)}
we know that $\text{tr}(\boldsymbol{c}_{0})=3$. Taking the trace
of equation \ref{eq:c0_eqn} allows us to solve for $a_{1}$ as a
function of $\boldsymbol{c}_{0}$ and $\nabla\boldsymbol{u}$

\[
\partial_{t}\text{tr}(\boldsymbol{c}_{0})+\ub\cdot\nabla\text{tr}(\boldsymbol{c}_{0})-2\gradu:\boldsymbol{c}_{0}+\text{tr}(\boldsymbol{c}_{0}-\boldsymbol{\delta})+6(1-a_{0})\text{tr}(\boldsymbol{c}_{1}+\beta_{1}a_{0}(\boldsymbol{c}_{0}-\boldsymbol{\delta})+\cdots
\]

\begin{equation}
+\beta_{0}a_{1}(\boldsymbol{c}_{0}-\boldsymbol{\delta})+\beta_{0}a_{0}\boldsymbol{c}_{1})+6(-a_{1})(\text{tr}(\boldsymbol{c}_{0})+\beta_{0}a_{0}\text{tr}((\boldsymbol{c}_{0}-\boldsymbol{\delta}))=0
\end{equation}

Combining $a_{0}=1$ and equation \ref{eq:tr(c0)}, this simplifies
to:

\begin{equation}
-2\gradu:\boldsymbol{c}_{0}-18a_{1}=0
\end{equation}

Therefore, we can solve for $a_{1}$ as

\begin{equation}
a_{1}=-\frac{1}{9}\gradu:\boldsymbol{c}_{0}\label{eq:a1}
\end{equation}

With $a_{1}=-\frac{1}{9}\nabla u:\boldsymbol{c}_{0}$ and equation
\ref{eq:a1_defn}, we can also obtain $\text{tr}(\boldsymbol{c}_{1})$

\begin{equation}
\text{tr}(\boldsymbol{c}_{1})=-6a_{1}=\frac{2}{3}\gradu:\boldsymbol{c}_{0}\label{eq:tr(c1)}
\end{equation}

Note that equations \ref{eq:c0_eqn} and \ref{eq:a1} combined are
equivalent to the nRP model, equation \ref{eq:full_nRP_ndim} after
substituting $a_{0}=1$.

\subsubsection{\protect\label{subsec:First-correction}First correction}

Moving to the $\mathscr{O}(\epsilon^{1})$ terms in equation \ref{eq:regrouped},
we find:

\[
\partial_{t}\boldsymbol{c}_{1}+(\ub\cdot\nabla\boldsymbol{c}_{1}-\gradu{}^{T}\cdot\boldsymbol{c}_{1}-\boldsymbol{c}_{1}\cdot\gradu)\text{Wi}+\boldsymbol{c}_{1}+6(-a_{2})\left[\boldsymbol{c}_{0}+\beta_{0}a_{0}(\boldsymbol{c}_{0}-\boldsymbol{\delta})\right]+
\]

\[
+6(-a_{1})\left[\boldsymbol{c}_{1}+\beta_{1}a_{0}(\boldsymbol{c}_{0}-\boldsymbol{\delta})+\beta_{0}a_{1}(\boldsymbol{c}_{0}-\boldsymbol{\delta})+\beta_{0}a_{0}\boldsymbol{c}_{1}\right]+6(1-a_{0})[\boldsymbol{c}_{2}+\cdots
\]

\begin{equation}
\beta_{2}a_{0}(\boldsymbol{c}_{0}-\boldsymbol{\delta})+\beta_{1}a_{1}(\boldsymbol{c}_{0}-\boldsymbol{\delta})+\beta_{1}a_{0}\boldsymbol{c}_{1}+\beta_{0}a_{1}\boldsymbol{c}_{1}+\beta_{0}a_{2}(\boldsymbol{c}_{0}-\boldsymbol{\delta})+\beta_{0}a_{0}\boldsymbol{c}_{2}]=0\label{eq:c1_eqn}
\end{equation}

In equation \ref{eq:c1_eqn}, we know $a_{0}=1$ and $a_{1}=-\frac{1}{9}\gradu:\boldsymbol{c}_{0}$,
but we need additional information from the second correction term,
namely $a_{2}$. Fortunately, equation \ref{eq:tr(c1)} defines the
trace of $\cb_{1}$ as a function of $\boldsymbol{c}_{0}$ and $\nabla\boldsymbol{u}$.
Similar to the procedure when solving for $a_{1}$ in equation \ref{eq:c0_eqn},
we once again take the trace of equation \ref{eq:c1_eqn} and this
yields an expression for $a_{2}$ as a function of $a_{1}$, $\boldsymbol{c}_{1}$
and $\nabla\boldsymbol{u}$

\begin{equation}
\frac{\partial}{\partial t}\left(\text{tr}(\boldsymbol{c}_{1})\right)+\ub\cdot\nabla(\text{tr}(\boldsymbol{c}_{1}))-2\gradu:\boldsymbol{c}_{1}+\text{tr}(\boldsymbol{c}_{1})-18a_{2}-6a_{1}(\text{tr}(\boldsymbol{c}_{1})+\beta_{0}\text{tr}(\boldsymbol{c}_{1}))=0
\end{equation}

With $\text{tr}(\boldsymbol{c}_{1})=-6a_{1}$, we can rearrange to
obtain a closed expression for $a_{2}$:

\begin{equation}
a_{2}=-\frac{1}{3}\left(\frac{\partial}{\partial t}a_{1}+\ub\cdot\nabla a_{1}\right)-\frac{1}{9}\gradu:\boldsymbol{c}_{1}+\left[-\frac{1}{3}+2\left(1+\beta_{0}\right)a_{1}\right]a_{1}
\end{equation}

This process can be extended to additional higher order correction
terms, but we see no engineering value beyond the first correction
terms outlined here.

\subsubsection{\protect\label{subsec:Stretch-dependent-CCR-expansions}Stretch-dependent
CCR expansions}

Equations \ref{eq:full_nRP_ndim} and \ref{eq:full_nRP_ndim} presume
a known functional form for $\beta_{0}$ and $\beta_{1}$, but thus
far we have avoided specifying a functional form for the CCR coefficient.
Here, we provide expressions for two possible variations that appear
in the literature.

First, the original Rolie Poly model uses a constant value of $\beta$,
which for our purposes implies $\beta_{0}$ is a constant and $\beta_{1}=\beta_{2}=\cdots=0$
for all higher order terms. However, a constant value of $\beta$
has proven problematic for reversing flows and a so-called regularized
Ianniruberto-Marrucci (IM) correction has been proposed \cite{ianniruberto2001simple}:

\begin{equation}
\beta(Z,a)=\beta_{\infty}\tanh(\Lambda Z(1-a))
\end{equation}

where $\Lambda\gg1$ is a regularization parameter and $\beta_{\infty}$
is the limiting value of $\beta$ that appears for steady flows when
$\Lambda\to\infty$. If we can assume a regular expansion for $\beta$:

\begin{equation}
\beta=\beta_{\infty}f(Z,a)=\beta_{\infty}\tanh(\Lambda Z(1-a))=\beta_{\infty}\tanh(\Lambda\frac{1}{\epsilon}(1-a_{0}-\epsilon a_{1}-\epsilon^{2}a_{2}))=-\beta_{\infty}\tanh(\Lambda(a_{1}+\epsilon a_{2}))
\end{equation}

To arrange $\beta$ as a series expansion in terms of $\epsilon$,
we use Taylor expansion.

\begin{equation}
\beta=\beta_{0}+\epsilon\beta_{1}+\epsilon^{2}\beta_{2}+\cdots\thickapprox-\beta_{\infty}(\tanh(\Lambda a_{1})+\Lambda a_{2}\epsilon\cosh^{-2}(\Lambda a_{1})-\Lambda^{2}a_{2}^{2}\epsilon^{2}\tanh(\Lambda a_{1})\cosh^{-2}(\Lambda a_{1})+\cdots)\label{eq:beta_tanh}
\end{equation}

Therefore, we have:

\begin{equation}
\beta_{0}=-\beta_{\infty}\tanh(\Lambda a_{1})
\end{equation}

\begin{equation}
\beta_{1}=-\beta_{\infty}\Lambda a_{2}\cosh^{-2}(\Lambda a_{1})
\end{equation}

\begin{equation}
\beta_{2}=\beta_{\infty}\Lambda^{2}a_{2}^{2}\epsilon^{2}\tanh(\Lambda a_{1})\cosh^{-2}(\Lambda a_{1})
\end{equation}

In general, we will advise on using the IM corrected version of the
Rolie Poly equation in all of its variations, including those derived
in this work.

\subsubsection{\protect\label{subsec:Consolidated-asymptotic-equation}Consolidated
asymptotic equations}

All combined, the equations of the second-order non-stretching Rolie
Poly (nRP2) model with the regularized IM correction are given by:

\begin{equation}
\overset{\nabla}{\boldsymbol{c}}_{0}=-(\boldsymbol{c}_{0}-\boldsymbol{\delta})+6a_{1}(\boldsymbol{c}_{0}+\beta_{0}(\boldsymbol{c}_{0}-\boldsymbol{\delta}))
\end{equation}

\begin{equation}
\overset{\nabla}{\boldsymbol{c}}_{1}=-\boldsymbol{c}_{1}+6a_{2}(\boldsymbol{c}_{0}+\beta_{0}(\boldsymbol{c}_{0}-\boldsymbol{\delta}))+6a_{1}(\boldsymbol{c}_{1}+\beta_{1}(\boldsymbol{c}_{0}-\boldsymbol{\delta})+\beta_{0}(a_{1}\boldsymbol{c}_{0}+\boldsymbol{c}_{1}-a_{1}\boldsymbol{\delta}))
\end{equation}

\begin{equation}
a_{1}=-\frac{1}{9}\boldsymbol{c}_{0}:\nabla u\hspace{1cm}a_{2}=-\frac{1}{3}\left(\frac{\partial}{\partial t}a_{1}+u\cdot\nabla a_{1}\right)-\frac{1}{9}\nabla u:\boldsymbol{c}_{1}+\left[-\frac{1}{3}+2\left(1+\beta_{0}\right)a_{1}\right]a_{1}
\end{equation}

\begin{equation}
\beta_{0}=-\beta_{\infty}\tanh(\Lambda a_{1})\hspace{1cm}\beta_{1}=-\beta_{\infty}\Lambda a_{2}\cosh^{-2}(\Lambda a_{1})
\end{equation}

\begin{equation}
\sigma=\boldsymbol{c}_{0}+\frac{1}{Z}\boldsymbol{c}_{1}-\boldsymbol{\delta}
\end{equation}

These equations are slightly more complex to implement than the established
Rolie Poly and nRP models. Compared to the Rolie Poly model, the nRP2
model equations are less stiff for numerical solutions at high entanglement
numbers. Compared to the nRP model, the nRP2 model provides a more
complete physical description of chain stretching and has no imminent
need for regularization.

\subsection{\protect\label{sec:nRP2_results}nRP2 Results and Discussion}

In principle, the nRP2 model should provide more accurate description
of chain stretching for flows with (dimensionless) shear rates of
$\gdot\sim\mathscr{O}(1)$ in the limit of $Z\to\infty$, but its
practical value for modeling flows with finite entanglement numbers
and large shear rates $\gdot\sim\mathscr{O}(Z)$ can only be assessed
via direct numerical simulation of the model. Predictions of the nRP2
model in complex time-dependent and spatially-resolved flows have
been reported elsewhere \cite{chen2024finite}, focusing specifically
on flows for which the nRP model exhibits pathological behaviors due
to its neglect of chain stretching. In this section, we will instead
focus on benchmarking the nRP2 model performance for predicting experimentally
measurable quantities in steady simple shear flow. 

In our study of steady simple shear flows, we will explore a range
of Wi, from slow deformations that barely cause polymers to orient,
$\wi=10^{-2}$, to fast deformations that induce significant chain
stretching, $\wi=10^{3}Z$. To quantify the error between the full
RP model and the nRP2 approximation, we introduce an error parameter
$e_{xy}$, defined as the relative error between the shear stress
predicted by the RP model and the nRP2 model:

\[
e_{xy}=\frac{|\sigma_{xy}^{\text{RP}}-\sigma_{xy}^{\text{nRP2}}|}{\sigma_{xy}^{\text{RP}}}
\]

We choose this error measure because it focuses on an experimentally
measurable feature of the model predictions. To contextualize the
limitations of the nRP2 model, we introduce a second parameter, $\wmax$.
 $\text{Wi}_{\text{max}}$ is defined as the maximum Weissenberg number
for which the nRP2 model agrees with the RP model to within a specified
tolerance, $e_{xy}<0.1$.

For systems with high entanglement numbers, we expect the nRP2 model
and RP model predictions to be closely aligned. At lower shear rates
$\wi\ll Z$, the non-stretching assumption remains valid, as the polymer
chains are near their equilibrium stretch, and predictions should
agree to $\mathscr{O}(1/Z^{2})$. As the shear rate increases, we
expect the nRP2 model to maintain this alignment, effectively capturing
polymer dynamics due to the reintroduction of chain stretching. This
should ensure good agreement between the models at moderate shear
rates, particularly before the Weissenberg number approaches the entanglement
number $\text{Wi}\sim Z$. For higher shear rates $\wi\gg Z$ or lower
entanglement numbers $Z\sim5$ there are no a-prior guarantees regarding
model performance, but through coincidence and good fortune asymptotic
approximations will often provide qualitatively useful approximations
when extended beyond the domain of the original derivation. In such
cases, however, it is important to exercise caution and skepticism
when interpreting model predictions.

In Figure \ref{fig:Stress_vs_Wi-1}, we present the results for shear
stress and normal stress as a function of the Weissenberg number for
a  steady shear flow with an entanglement number $Z=1000$. As expected,
the model shows good agreement with the RP solution up to $\text{Wi}\sim Z$.
However, beyond this threshold, the nRP2 model begins to deviate,
with the vertical dotted line identifying $\wi_{\text{max}}$, i.e.
$e_{xy}>0.1$. This deviation highlights the limits of the nRP2 model
in capturing the full range of polymer dynamics at high shear rates. 

{\bfseries{}
\begin{figure}
\centering{}\includegraphics[scale=0.25]{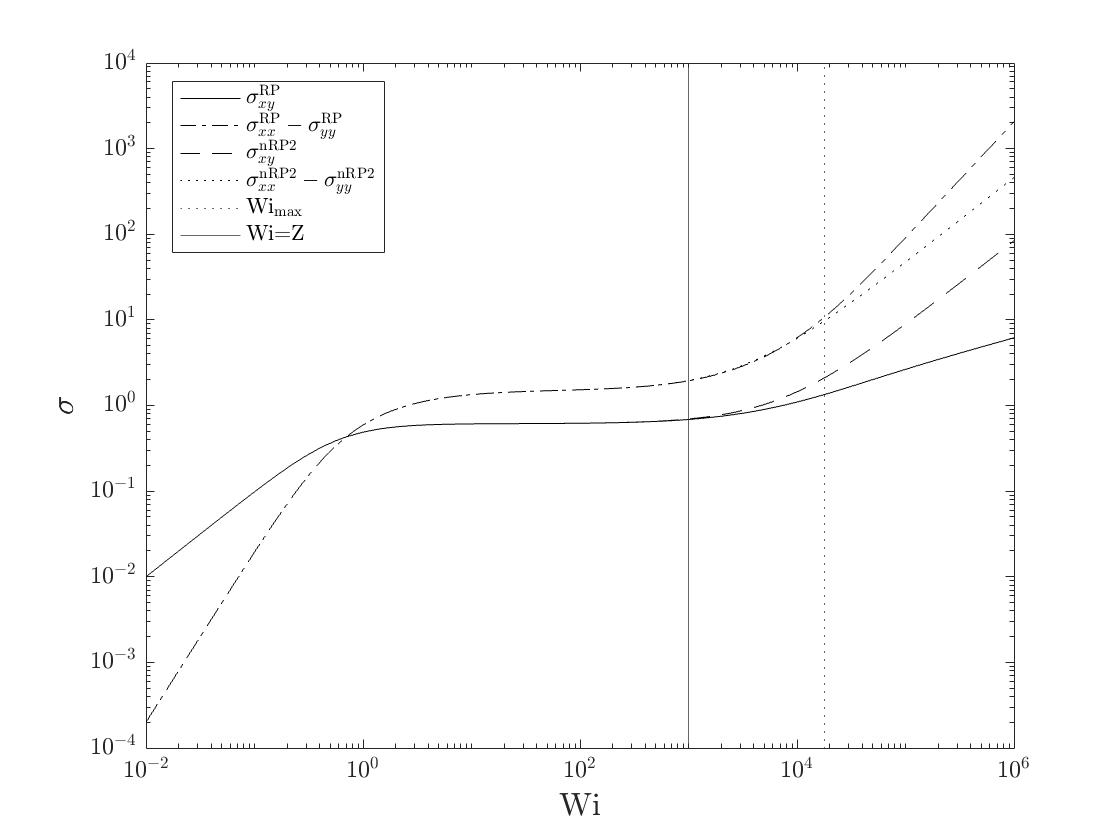}\caption{\protect\label{fig:Stress_vs_Wi-1}Shear stress vs $\protect\wi$
for a steady shear flow with $Z=1000$. For $\protect\wi<Z$, the
nRP2 model shows a good alignment with RP predictions. The vertical
dotted line indicates a cutoff, as the shear stress starts to deviate
by more than 10\%.}
\end{figure}
}

\begin{figure}
\begin{centering}
\includegraphics[scale=0.25]{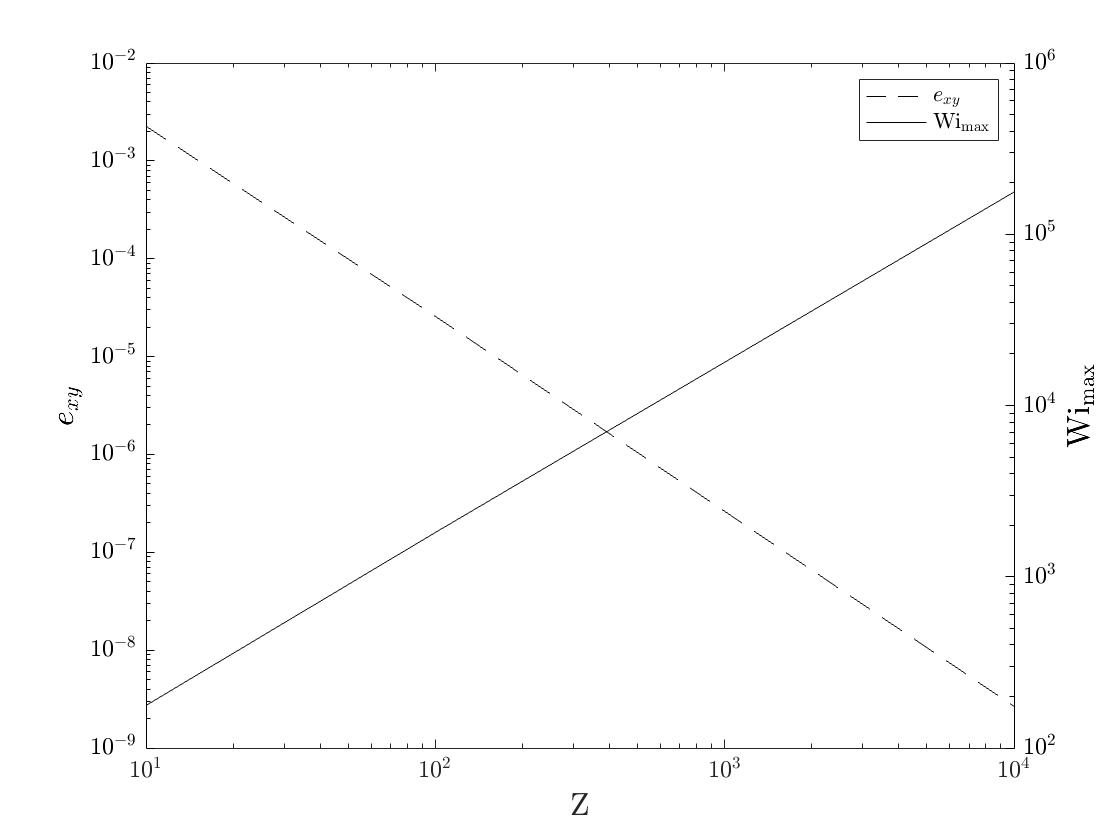}
\par\end{centering}
\caption{\protect\label{fig:Correlation-1}Deviations between RP and nRP2 predictions
behave predictably. For a fixed $\protect\wi<Z$, in this case $\protect\wi=3$,
we see that the error $e_{xy}$ scales as $e_{xy}\sim Z^{-2}$. For
a fixed tolerance $e_{xy}$, we likewise see that that the maximum
range of the nRP2 model scales as $\protect\wi_{\text{max}}\sim Z$.}
\end{figure}

Next, we will attempt to quantify the error of the nRP2 model when
it is employed in its range of intended use, $\wi\ll Z$. In Figure
\ref{fig:Correlation-1}, we plot the error $e_{xy}$ of the nRP2
model as a function of the entanglement number $Z$. The dashed line
(left vertical axis) gives the relationship between $e_{xy}$ and
$Z$ at a fixed Wi = 3, revealing a clear scaling of $e_{xy}\sim Z^{-2}$
indicating agreement up to the order of $Z^{-1}$ as anticipated from
the model derivation. The solid line in Figure \ref{fig:Correlation-1}
(right vertical axis) shows the proportionality between $\text{Wi}_{\text{max}}$
and $Z$, indicating that higher entanglement numbers enable the model
to maintain accuracy over a broader range of Weissenberg numbers,
while maintaining the same limiting scaling of $\wmax\sim Z$.

In summary, the nRP2 model successfully reintroduces chain stretching
into the non-stretching RP framework, addressing the limitations of
the non-stretching approximation, particularly at higher deformation
rates where chain stretching becomes significant. We validated the
model's effectiveness by testing its predictions on the shear stress
and normal stress with varying entanglement numbers. Our simulations
demonstrate that the nRP2 model closely aligns with the RP model at
low to moderate shear rates for highly entangled systems, accurately
capturing corrections to polymer dynamics incurred by chain stretching.
However, as the Weissenberg number approaches the entanglement number,
$\wmax\sim Z$, deviations begin to emerge as chain stretching is
no longer a perturbative correction to the polymer's relaxation behavior.
For systems with low entanglement numbers, $Z<10$, the nRP2 model
does not provide a quantitatively useful approximation for the RP
model.

In the supplemental materials, we have included MATLAB code that extends
the analysis of this section to extensional flows---uniaxial, biaxial,
and planar---where the same trends of error scaling $\mathscr{O}(Z^{-2})$
and model range $\wmax\sim Z$ are preserved. This suggests that the
strengths and limitations of the nRP2 model are independent of flow
type, and the analysis for simple shear flow reported here is representative.

\section{\protect\label{sec:nDO}Entanglement Dynamics}

\subsection{\protect\label{subsec:DO_starting_eqns}Starting Equations}

The RP model, following its predecessor models \cite{likhtman2000microscopic,graham2003microscopic,milner2001microscopic},
does not account for the possibility of a tube diameter or entanglement
number that is significantly modified by flow. Recently, Dolata and
Olmsted (DO) introduced a dynamical evolution equation for chain entanglement
density $\nu$ with thermodynamically consistent couplings to stress
and stress relaxation \cite{dolata2023thermodynamically}. Polymer
relaxation processes were based on the RP model, and  incorporated
the tube-model informed kinetic equation from the Ianniruberto-Marrucci
to capture flow-induced disentanglement \cite{ianniruberto2014convective}.
Translating the DO model into the notation of the current manuscript,
a configuration tensors $\cb$  describes the stress from chains with
equilbrium entanglement number $\zeq$ that, under flow, takes on
a non-equilibrium value  $Z=\nu\zeq$. As entanglements are released,
conformational relaxation occurs, leading to a decrease in the effective
reptation time. To account for this, we use the dimensionless effective
reptation time $\tau_{D}(\lambda,\beta)$, defining as the effective
reptation time divided by the equilibrium reptation time. As with
the RP model, we scale time by the reptation time at equilibrium $\tau_{D\text{eq}}$.
Following Dolata and Olmsted, we choose the re-entanglement time $\tau_{\nu}$
 to be identical to the Rouse time \cite{o2019stress}, and thus in
the dimensionless expression $\tau_{D\text{eq}}/\tau_{\nu}=3\zeq$. 

The DO model also allows for finite extensibility corrections and
Giesekus-like terms, but these will not be featured in our analysis.
Instead, we focus on the author's equations IV.38 for a better direct
comparison against the nRP model, to understand the effect of entanglement
dynamics specifically. Overall, the nondimensionalized DO model can
be written as:

\begin{equation}
\overset{\nabla}{\cb}=-(\cb-\db)-6\zeq(1-a)\left[\cb+\left(\frac{\zeq\nu}{\zeq\nu+1}\right)\frac{\beta}{2}(\cb-\db)\right]-\frac{\zeta_{z}\beta\nu}{3}a^{2}\left[\frac{1}{\tau_{D}(a,\beta)}+\frac{6\zeq}{(1+1/a)}\right]\cb\ln\nu
\end{equation}

\begin{equation}
\zeta_{z}=\frac{2}{1+1/\zeq}
\end{equation}

\begin{equation}
\frac{1}{\tau_{D}(\lambda,\beta)}=1+3\zeq\left(\frac{\zeq\nu}{\zeq\nu+1}\right)\beta(1-a)
\end{equation}

\begin{equation}
\frac{D\nu}{Dt}=-\frac{\beta\nu}{3}a^{2}\left(\cb:\nabla\nu-\frac{1}{2}\frac{D\text{tr}\cb}{Dt}\right)-3\zeq\ln\nu
\end{equation}

\begin{equation}
\sb=\cb-\db
\end{equation}

\subsection{\protect\label{subsec:DO_Derivation}Derivation}

We will pursue a leading-order description of the DO model in its
"non stretching" limit, i.e. highly entangled flows $\zeq\gg1$
at shear rates $\wi\ll\zeq$. Assuming a regular perturbation expansion
for all configuration tensors, $\cb=\cb_{0}+\zeq^{-1}\cb_{1}+\mathscr{O}(\zeq^{-2})$
we follow the same procedure as outlined in the preceding section.
The kinetic equation for disentanglement is similarly expanded, $\nu=\nu_{0}+\zeq^{-1}\nu_{1}+\mathscr{O}(\zeq^{-2})$.

In the $\mathscr{O}(\zeq)$ equations, we once again find that chains
must be non-stretching to leading order, and we find the entanglement
density $\nu$ is likewise equilibrated.

\begin{equation}
0=0-6\zeq(1-a_{0})\left[\cb_{0}+\frac{\zeq\nu}{\zeq\nu+1}\frac{\beta}{2}(\cb_{0}-\db)\right]-\frac{\zeta_{z}\beta\nu_{0}}{3}a_{0}^{2}\left[\frac{1}{\tau_{D}(a,\beta)}+\frac{6\zeq}{(1+1/a)}\right]\cb\ln\nu_{0}
\end{equation}

\begin{equation}
0=0-3\zeq\ln\nu_{0}
\end{equation}

\begin{equation}
a_{0}=1
\end{equation}

\begin{equation}
\nu_{0}=1
\end{equation}

Proceeding to the $\mathscr{O}(1)$ equations, we find:

\begin{equation}
\overset{\nabla}{\cb}_{0}=-(\cb_{0}-\db)+6a_{1}\left[\cb_{0}+\frac{\beta}{2}(\cb_{0}-\db)\right]-2\beta\cb_{0}\nu_{1}
\end{equation}

\begin{equation}
0=-\frac{\beta}{3}(\cb_{0}:\nabla u)-3\nu_{1}
\end{equation}

However, since all chains are non-stretched, $a_{0}=1$, it must be
that $\text{tr}\cb=3$ at all times. Taking the trace of the configuration
equation and rearranging  we find:

\begin{equation}
a_{1}=-\frac{1}{9}\gradu:\boldsymbol{c}_{0}+\frac{1}{3}\beta\nu_{1}
\end{equation}

\begin{equation}
\nu_{1}=-\frac{\beta}{9}(\cb_{0}:\gradu)
\end{equation}

Overall, the leading-order solution for the non-stretching Dolata-Olmsted
(nDO) model is given by:

\begin{equation}
\overset{\nabla}{\cb}=-(\cb-\db)-\frac{2}{3}(\cb_{0}:\gradu)(\cb+\beta^{\text{eff}}(\cb-\db))
\end{equation}

\begin{equation}
\beta^{\text{eff}}=\frac{1}{2}\left[1-\frac{1}{3}\beta^{2}\right]\beta
\end{equation}

\begin{equation}
\nu=1-\frac{1}{\zeq}\frac{\beta}{9}(\cb_{0}:\nabla u)+\mathscr{O}(1/\zeq^{2})
\end{equation}

Here, we see that the nDO model is identical to the nRP model, but
with a reinterpretation of the CCR coefficient $\beta^{\text{eff}}$.
In some sense, this should not be surprising; the non-disentangled
limit is equivalent to the non-stretching limit if $\tau_{\nu}=\tau_{R}$
as the authors recommend, so there is no obvious mechanism by which
the models might fundamentally differ under such circumstances. For
a comparison of the DO and nDO model predictions, we refer the reader
to the supplemental materials section, which contains a working Matlab
code for a simple shear flow analyis in both models.

\section{\protect\label{sec:nRDP}Polydispersity}

\subsection{\protect\label{subsec:nRDP_Starting-Equations}Starting Equations}

To extend the non-stretching approximation to highly entangled polydisperse
melts of linear polymers, we will use the Rolie Double Poly (RDP)
model \cite{boudara2019nonlinear}. Given a discrete molecular weight
distribution with $n$ distinct polymer molecular weights $M_{1},M_{2},...,M_{n}$
with volume fractions $\phi_{1},\phi_{2},...,\phi_{n}$, we construct
configuration tensors $\cb_{ij}$ to describe the stress from chains
of molecular weight $M_{i}$ arising due to their entanglements with
chains of molecular weight $M_{j}$. Scaling time by the reptation
time for a chain of molecular weight $M_{W}=\sum_{i=1}^{n}\phi_{i}M_{i}$,
and assuming $\tau_{Di}\sim M_{i}^{3}$ and $\tau_{Ri}\sim M_{i}^{2}$,
the configuration tensors $\cb_{ij}$ evolve by:

\begin{equation}
\overset{\nabla}{\cb}_{ij}=-\left[\frac{1}{z_{i}^{3}}+\frac{1}{z_{j}^{3}}\right](\cb_{ij}-\db)-6\bar{Z}(1-a_{i})\frac{1}{z_{i}^{2}}\cb_{ij}-6\bar{Z}(1-a_{j})\frac{1}{z_{j}^{2}}a_{i}\beta_{i}(\cb_{ij}-\db)\label{eq:RDP_ndim}
\end{equation}

\begin{equation}
a_{i}=\frac{1}{\lambda_{i}}=\sqrt{\frac{3}{\text{tr}\cb_{i}}}\hspace{1cm}\cb_{i}=\sum_{j=1}^{n}\phi_{j}\cb_{ij}
\end{equation}

\begin{equation}
\beta_{i}=\beta\tanh(\Lambda\bar{Z}(1-a_{i})z_{i}^{-2})
\end{equation}

\begin{equation}
\sb=\left[\sum_{i=1}^{n}\phi_{i}\cb_{i}\right]-\db
\end{equation}

In equation \ref{eq:RDP_ndim}, we define $z_{i}=M_{i}/M_{W}$ as
a rescaled chain length, and $\bar{Z}=M_{W}/M_{e}$ as the entanglement
number for a chain with the system's weight-average molecular weight.
As in section \ref{sec:Equations-and-Nondimensionalizat}, we assume
for simplicity that $\tau_{D,i}/\tau_{R,i}=3Z_{i}$ for all chains.
To be consistent with the RDP's microscopic interpretation of the
configuration tensors $\cb_{ij}$, the regularized IM correction to
the CCR coefficient $\beta_{i}$is a function of the rate of chain
retraction in chain $i$ species rather than the $j$ species. For
systems characterized by broad-spectrum (as opposed to discrete) polydispersity,
the continuous molecular weight distribution can be approximated as
a discrete spectrum by binning chains into discrete intervals \cite{carrot1997dynamic}.

\subsection{\protect\label{subsec:nRDP_derivation}Derivation}

We will pursue a leading-order description of the RDP model in its
"non stretching" limit, i.e. highly entangled flows $\bar{Z}\gg1$
at shear rates $\wi\ll\bar{Z}$. Assuming a regular perturbation expansion
for all configuration tensors, $\cb_{ij}=\cb_{ij,0}+\bar{Z}^{-1}\cb_{ij,1}+\mathscr{O}(\bar{Z}^{-2})$
we follow the same procedure as outlined in subsections \ref{subsec:Setting-up-Asymptotic}
- \ref{subsec:Leading-order-solution}.

In the $\mathscr{O}(\bar{Z})$ equations, we once again find that
chains must be non-stretching to leading order:

\begin{equation}
0=-6(1-a_{i,0})\frac{1}{z_{i}^{2}}\cb_{ij,0}-6(1-a_{j,0})\frac{1}{z_{j}^{2}}a_{i,0}\beta_{i,0}(\cb_{ij,0}-\db)
\end{equation}

\begin{equation}
a_{1,0}=a_{2,0}=\cdots=a_{n,0}=1
\end{equation}

Proceeding to the $\mathscr{O}(1)$ equations, we find:

\begin{equation}
\overset{\nabla}{\cb}_{ij,0}=-\left[\frac{1}{z_{i}^{3}}+\frac{1}{z_{j}^{3}}\right](\cb_{ij,0}-\db)+6a_{i,1}\frac{1}{z_{i}^{2}}\cb_{ij,0}+6a_{j,1}\frac{1}{z_{j}^{2}}\beta_{i,0}(\cb_{ij,0}-\db)\label{eq:nRDP_order1}
\end{equation}

However, since all chains are non-stretching to leading order, $a_{i,0}=1$,
it must be that $\text{tr}\cb_{i,0}=\sum_{k=1}^{n}\phi_{k}\text{tr}\cb_{ik,0}=3$
at all times. Taking the trace of equation \ref{eq:nRDP_order1} and
summing over all species, we find:

\begin{equation}
-2\cb_{i}:\gradu=-\left[\sum_{k=1}^{n}\phi_{k}\left[\frac{1}{z_{k}^{3}}-\frac{6}{z_{k}^{2}}\beta_{i,0}a_{k,1}\right](\text{tr}\cb_{ik,0}-3)\right]+18a_{i,1}\frac{1}{z_{i}^{2}}\label{eq:nRDP_a1}
\end{equation}

Applying equation \ref{eq:nRDP_a1} to all $n$ chain indices yields
a system of $n$ equations and $n$ unknown values of $a_{i,0}$ that
can be solved numerically. Perhaps somewhat unexpectedly, the non-stretching
approximation as applied to the RDP model does not yield $\text{tr}\cb_{ij,0}=3$.
Instead, we only find that the overall configuration tensor for any
chain shows no net stretching, $\text{tr}\cb_{i,0}=3$, to leading
order. We do not comment on whether this prediction is physically
meaningful, we only point out that it is a feature of the RDP model.

Overall, the leading-order solution for the non-stretching RDP (nRDP)
model is given by:

\begin{equation}
\overset{\nabla}{\cb}_{ij,0}=-\left[\frac{1}{z_{i}^{3}}+\frac{1}{z_{j}^{3}}\right](\cb_{ij,0}-\db)+6a_{i,1}\frac{1}{z_{i}^{2}}\cb_{ij,0}+6a_{j,1}\frac{1}{z_{j}^{2}}\beta_{i,0}(\cb_{ij,0}-\db)\label{eq:nRDP_start}
\end{equation}

\begin{equation}
-2\cb_{i}:\gradu=-\left[\sum_{k=1}^{n}\phi_{k}\left[\frac{1}{z_{k}^{3}}-\frac{6}{z_{k}^{2}}\beta_{i,0}a_{k,1}\right](\text{tr}\cb_{ik,0}-3)\right]+18a_{i,1}\frac{1}{z_{i}^{2}}
\end{equation}

\begin{equation}
\cb_{i,0}=\sum_{j=1}^{n}\phi_{j}\cb_{ij,0}
\end{equation}

\begin{equation}
\text{tr}\cb_{i,0}=3
\end{equation}

\begin{equation}
\beta_{i,0}=\beta\tanh(-\Lambda\bar{Z}a_{i,0}z_{i}^{-2})
\end{equation}

\begin{equation}
\sb=\left[\sum_{i=1}^{n}\phi_{i}\cb_{i,0}\right]-\db
\end{equation}

For a comparison of the RDP and nRDP model predictions, we refer the
reader to the supplemental materials section, which contains a working
Matlab code for start-up simple shear flow in both models.

\section{\protect\label{sec:nLRP}Polymer Reactions}

\subsection{\protect\label{subsec:LRP_starting_equations}Starting Equations}

In systems of well-entangled living polymers, reversible scission
reactions speed up stress relaxation and produce a narrower stress
relaxation spectra. Because living polymers are well approximated
as having a continuous molecular weight distribution, we will describe
a configuration tensor $\cb(z)$ as a function of a dimensionless
chain length $z=L/\bar{L}$ where $\bar{L}$ is the number-average
molecular weight of the distribution $n(L)=n_{o}e^{-L/\bar{L}}$.
Living polymer systems also have an additional dimensionless number,
$\zeta$, which compares the typical time for a chain to break (or
otherwise rearrange its monomer sequence) to the equilibrium stress
relaxation time if breaking were suppressed.

Constitutive models for living polymer rheology can be exceedingly
complex, but for illustration purposes we will focus on a variation
of the "Living Rolie Poly" model \cite{cates1987reptation} in
which the (fully equilibrated) reversible scission reactions are replaced
by a simpler "shuffling" approximation \cite{peterson2021predictions}.

\begin{equation}
\overset{\nabla}{\cb}(z)=-\frac{1}{z^{3}}(\cb-\db)-6\bar{Z}\frac{1}{z^{2}}(1-a)(\cb+\beta a(\cb-\db))-\frac{1}{\zeta}(\cb-\bar{\cb})
\end{equation}

\begin{equation}
a(z)=\sqrt{\frac{3}{\text{tr}\cb(z)}}
\end{equation}

\begin{equation}
\beta(z)=\bar{\beta}\tanh(\bar{Z}\Lambda(1-a(z)z^{-2})
\end{equation}

\begin{equation}
\bar{\cb}=\int_{0}^{\infty}dze^{-z}\cb(z)
\end{equation}

\begin{equation}
\sb=\bar{\cb}-\db
\end{equation}

\subsection{\protect\label{subsec:nLRP_derivation}Derivation}

Unlike the RDP model, the LRP model uses a "single reptation"
description of all stress relaxation processes - there are no nonlinear
relaxation couplings between different sectors of the molecular weight
distribution. Instead, the coupling terms that emerge via the shuffling
mechanism are strictly linear in $\cb(z)$. The same would be true
for any linear mixing process, from reversible scission to end attack
or bond interchange - this is not an artifact unique to the shuffling
approximation. Due to the lack of nonlinear relaxation couplings,
the non stretching approximation of the LRP model trivially reduces
the form that one might anticipate directly from the nRP model.

We will pursue a leading-order description of the LRP shuffling model
in its "non stretching" limit, i.e. highly entangled flows $\bar{Z}\gg1$
at shear rates $\wi\ll\bar{Z}$. Assuming a regular perturbation
expansion for all configuration tensors, $\cb(z)=\cb_{0}(z)+\bar{Z}^{-1}\cb_{1}(z)+\mathscr{O}(\bar{Z}^{-2})$
we follow the same procedure as outlined in subsections \ref{subsec:Setting-up-Asymptotic}
- \ref{subsec:Leading-order-solution}.

In the $\mathscr{O}(\bar{Z})$ equations, we once again find that
chains must be non-stretching to leading order:

\begin{equation}
0=\frac{1}{z^{2}}(1-a_{0})(\cb_{0}+\beta_{0}a_{0}(\cb_{0}-\db))
\end{equation}

\begin{equation}
a_{0}(z)=1
\end{equation}

Proceeding to the $\mathscr{O}(1)$ equations, we find:

\begin{equation}
\overset{\nabla}{\cb}_{0}(z)=-\frac{1}{z^{3}}(\cb_{0}-\db)+6\bar{Z}\frac{1}{z^{2}}a_{1}(\cb_{0}+\beta_{0}(\cb_{0}-\db))-\frac{1}{\zeta}(\cb_{0}-\bar{\cb}_{0})\label{eq:nLRP_order1}
\end{equation}

However, since all chains are non-stretched, $a_{0}(z)=1$, it must
be that $\text{tr}\cb(z)=3$ at all times. Taking the trace of equation
\ref{eq:nLRP_order1} and summing over all species, we find:

\begin{equation}
a_{1}(z)=-\frac{z^{2}}{9}\cb(z):\gradu
\end{equation}

Inserting this expression for $a_{1}$ back into equation \ref{eq:nLRP_order1},
we see that the non stretching LRP (nLRP) model looks like the standard
nRP model with a shuffling term added on:

\begin{equation}
\overset{\nabla}{\cb}_{0}(z)=-\frac{1}{z^{3}}(\cb_{0}-\db)-\frac{2}{3}\left[\cb:\gradu\right](\cb_{0}+\beta_{0}(\cb_{0}-\db))-\frac{1}{\zeta}(\cb_{0}-\bar{\cb}_{0})
\end{equation}

\begin{equation}
\bar{\cb}_{0}=\int_{0}^{\infty}dze^{-z}\cb(z)
\end{equation}

\begin{equation}
\text{tr}\cb(z)=3
\end{equation}

\section{\protect\label{sec:supplemental}Description of Supplemental Materials}

The Supplemental Information (SI) provides MATLAB code that enables
reproduction of the key figures and results presented in Sections
2, 3, and 4 of this manuscript. The SI includes:
\begin{itemize}
\item A set of MATLAB scripts to reproduce the figures from section 2.3
(section \ref{sec:nRP2_results}), including a main script for promblem
setup and two function scripts for both the RP and nRP2 models. These
codes enable replicating Figure 1 (\ref{fig:Stress_vs_Wi-1}) and
Figure 2 (\ref{fig:Correlation-1}). Additionally, we include the
code that extends the similar analysis with the extensional flows--uniaxial,
biaxial, and planar flows.
\item A set of scripts comparing the predictions of nDO and DO models, as
discussed in sections 3 (section \ref{sec:nDO}). A main script setting
up a simple shear flow and two function scripts for DO and nDO models
are included. 
\item A set of scripts for the comparison of RDP and nRDP models, as in
section 4 (section \ref{sec:nRDP}). These scripts enables a start-up
flow simulation, with a problem setup script, two function scripts
for RDP and RDP, and a function script for the molecular weight distribution.
\end{itemize}
These codes perform as expected, and to our estimation the figures
do not yield any surprising results compared to what we discuss in
section 2.3 (section \ref{sec:nRP2_results}). For this reason, we
do not include more visualized results in the manuscript. 

\section{\protect\label{sec:Conclusions}Conclusions}

In this study, we have derived and validated non-stretching approximations
for the RP model and some of its successor variants. We improved the
non-stretching version of the RP model by extending it to second order
in accuracy, reintroducing chain stretching. We also extended our
analysis to modified RP models that account for additional physics
such as flow-induced disentanglement, polydispersity, and reversible
scission reactions. In each case, the analysis was conducted by constructing
an asymptotic solution to the governing equations in the limit of
high entanglement $Z\gg1$ and fixed Weissenberg number $\wi\ll Z$.
Results were obtained to second order for the RP model and first order
for all of the subsequent variants considered.

Although slightly more complex to implement, the nRP2 model provides
a more accurate physical description of chain stretching without introducing
non-physical means of regularization. Our simulations validate the
model's accuracy in predicting shear and normal stresses across various
entanglement numbers $Z\gg1$, demonstrating good alignment with the
RP model for dimensionless shear rates that are still not strongly
stretching, $\wi<Z$. When applying the non-stretching approximation
to Rolie Poly variants, the non-stretching model performed as expected,
similar to what has previously been shown for nRP vs RP.

However, we acknowledge limitations in our study. Deviations between
the full models and their non-stretching variants are consistently
seen at high dimensionless shear rates, $\wi>Z$, and in weakly entangled
systems, $Z\sim1$, reflecting a breakdown of the assumptions under
which the model was derived. Quantitative improvements to the non-stretching
approximation can be achieved by proceeding to higher-order approximations,
but we do not foresee a significant engineering benefit to this approach
as it will not necessarily improve the model accuracy for $\wi>Z$
flows. Likewise, for $\wi>Z$ flows where chain stretching is dominant,
one could derive an analagous "non-reptating" model, but the potential
engineering benefits of having such a model are not clear. 

In the future, we hope to implement the nRP2 model in an open source
computational fluid dynamics package such as rheoTool to test its
accuracy and numerical efficacy in complex flow conditions. The nRP2
model will also be useful for studying the effects of chain stretching
in LAOS flow to make it easier to study the effects of chain stretching
at varying entanglement numbers.
\printbibliography
\end{document}